\documentclass[a4paper,11pt]{article}
\pdfoutput=1 

\usepackage{jheppub} 

\usepackage[T1]{fontenc} 
\usepackage{float}
\usepackage{epstopdf}
\usepackage{amsmath}
\makeatletter
\gdef\@fpheader{}
\makeatother

\title{\boldmath Power-Yang-Mills black holes and black branes in quartic quasi-topological gravity}

\author[1,2]{Askar Ali}

\author[1,3]{and Khalid Saifullah}

\affiliation[1]{Department of Mathematics, Quaid-i-Azam University, Islamabad, Pakistan}

\affiliation[2]{Department of Sciences and Humanities, National University of Computer and Emerging Sciences, Peshawar 25000, Pakistan}

\affiliation[3]{School of Mathematical Sciences, Queen Mary University of London, London, United Kingdom}

\emailAdd{askarali@math.qau.edu.pk}\emailAdd{ksaifullah@fas.harvard.edu}

\abstract{We study higher dimensional quartic quasi-topological black holes in the framework of non-abelian power-Yang-Mills theory. It is shown that real solutions of the gravitational field equations exist only for positive values of quartic quasi-topological coefficient. Depending on the values of the mass parameter and Yang-Mills charge, they can be interpreted as black holes with one horizon, two horizons and naked singularity. It is also shown that the solution associated with these black holes has an essential curvature singularity at the centre $r=0$. Thermodynamic and conserved quantities for these black holes are computed and we show that the first law has been verified. We also check thermodynamic stability in both canonical and grand canonical ensembles. In addition to this, we also formulate new power-Yang-Mills black hole solutions in pure quasi-topological gravity. The physical and thermodynamic properties of these black holes are discussed as well. It is concluded that unlike Yang-Mills black holes there exist stability regions for smaller power-Yang-Mills black holes in grand canonical ensemble. Finally, we discuss the thermodynamics of horizon flat power-Yang-Mills rotating black branes and analyze their thermodynamic and conserved quantities by using the counter-term method inspired by AdS/CFT correspondence.       
\vspace{60 mm}
}

\notoc 

\begin{document}
\maketitle
\flushbottom 


\section{Motivation}
\label{sec:intro}
Higher dimensional gravities sometimes give rise to more interesting possibilities than the four-dimensional theories. This major leap can help solve the problem of ``hierarchy of scales''. Many higher dimensional models have been formulated in recent years. The well-known generalization of Einstein's gravity is the Lovelock gravity \cite{1W,2W,3W}. The equations of motion obtained in this gravity are still second order. Due to the topological background of Lovelock gravity, the corresponding Gauss-Bonnet term in the action does not possess any dynamical contribution in four dimensional geometries. Similarly, the third order Lovelock term gives contributions to the gravitational field equations in spacetime dimensions greater than or equal to seven. The generalization to this theory which contains the cubic and quartic curvature terms and possess dynamical contributions in five spacetime dimensions is the quasi-topological gravity \cite{4W,5W,6W}. The equations of motion for second, third \cite{4W,5W}, and fourth order \cite{6W} quasi-topological gravities are also second order for the spherically symmetric metric and are valid in five and higher dimensions. Because quasi-topological gravity does not contain derivatives of metric more than second order (for spherically symmetric case), the quantization of its linearized form is ghost-free.

The holographic study of four and higher dimensional conformal field theories could also be possible in the background of quasi-topological gravity \cite{5W}. The lower non-zero value in a specific corner of the possible space of gravitational couplings for the ratio of the shear viscosity and entropy can also be found in this theory \cite{7W}. In AdS/CFT correspondence, quasi-topological gravity can provide enough free coupling parameters which make a one-to-one relationship with the central charges and in this way gravitational spacetimes can be constructed \cite{5W,8W,9W,10W,11W,12W}. Furthermore, since the terms of quasi-topological gravity are not truly topological, so, the nontrivial gravitational effects in fewer dimensions are also possible. Therefore, this theory has priority over the Lovelock gravity \cite{6W}. By imposing particular constraints on the coupling parameters of this modified gravity, the causality for the CFT can be respected \cite{13W,14W,15W}. Thus, it might be very interesting to study black holes and black branes in quasi-topological gravity. In this context, quasi-topological black holes have been investigated in the literature \cite{4W,5W,16W,17W,18W}. Two families of solutions for the neutral and Maxwellian charged quasi-topological black holes were derived in Refs. \cite{6W,19W}. The black hole solution in cubic quasi-topological gravity with power-Maxwell source has been constructed in Ref. \cite{20S}. The solutions describing Lifshitz quartic quasi-topological black holes were found in Ref. \cite{20W}. 

The modification of abelian Maxwell's theory is the non-abelian Yang-Mills theory which is used in the study of gauged AdS super-gravity theories. The assumption of Yang-Mills field as a source of gravity was given in Ref. \cite{21W}. Under this assumption, a class of asymptotically flat spherically symmetric Yang-Mills solutions was derived numerically. The black holes of Einstein's gravity with this source have been studied in Ref. \cite{22W}. Similarly, Yang-Mills black holes with cosmological constant in Einstein's theory were studied in Refs. \cite{23W,24W,25W}. Other investigations of black holes within this framework of Yang-Mills field are in Refs. \cite{26W,27W,28W,29W}. A new family of black holes in Lovelock-Yang-Mills theory were also introduced in Ref. \cite{29W}. By taking the Wu-Yang ansatz \cite{30W}, the authors of Ref. \cite{31W} derived the analytical solution describing black holes in quasi-topological-Yang-Mills gravity. Instead of Yang-Mills theory, one can also couple power-Yang-Mills theory with gravity and explore black holes \cite{32W}, i.e., to consider the source as $(F^{(a)}_{\alpha\beta}F^{(a)\alpha\beta})^q$, where $F^{(a)}_{\alpha\beta}$ is the Yang-Mills field with $1\leq a\leq (d-1)(d-2)/2$ and $q$ is a parameter of nonlinearity. Using this idea, the black holes of Lovelock gravity were studied and new third order Lovelock as well as Gauss-Bonnet solutions were found \cite{32W}. Similarly, dimensionally continued power-Yang-Mills black holes \cite{33W} and Lovelock-power-Yang-Mills black holes surrounded by dark fluid \cite{34W} have also been found recently. In this paper, we are taking the Wu-Yang ansatz \cite{30W} for the study of power-Yang-Mills black holes in quartic quasi-topological gravity. 

The physics of black holes in pure Lovelock gravity has attracted much attention from theoretical physicists. Recently, black holes of this theory with different matter sources have been studied in the literature \cite{35W,36W,37W,38W,AS22}. It is shown in Ref. \cite{39W} that the black hole of $d=3N+1$ dimensional pure Lovelock gravity is stable. A study related to the ADM mass and quasi-local energy in this theory is presented in Ref. \cite{40W}. Similarly, thermodynamic behaviour and PV criticality of pure Lovelock black holes were also investigated \cite{41W}. In addition to the Yang-Mills solution representing black holes of quartic quasi-topological gravity, the authors of Ref. \cite{31W} derived the Yang-Mills black hole solution in pure quasi-topological theory as well. Motivated by this work, we discuss power-Yang-Mills black holes of pure quasi-topological theory in this paper. 

Recently, rotating black branes in Einstein's theory with nonlinear electromagnetic sources have been studied \cite{47W}. The generalization of these nonlinearly charged rotating black branes in Gauss-Bonnet gravity have also been worked out \cite{48W}. Furthermore, thermodynamics of rotating Lovelock black branes with Maxwell \cite{49W,49H} and nonlinear electromagnetic sources \cite{50W,51W,52W} has also been probed. Similarly, the rotating black branes of quasi-topological gravity and their thermodynamic properties have been studied \cite{53W,54W}. In this paper, we investigate black branes of quartic quasi-topological gravity when coupled with the power-Yang-Mills field. A new class of Yang-Mills black branes will also be recovered from our results when we put $q=1$.
  
   The outline of this paper is as follows. In Section 2, we construct the action function associated with $d$-dimensional quartic quasi-topological gravity and use the model of power-Yang-Mills theory for the determination of new black hole solutions. In Section 3, we investigate the thermodynamic properties and validity of the first law for these objects. In Section 4, we work out the physical properties of power-Yang-Mills black holes in pure quasi-topological gravity. Thermodynamic stability of these pure quasi-topological black holes are discussed in Section 5. Further, Section 6 is devoted to the thermodynamic properties of power-Yang-Mills rotating black branes and their associated conserved quantities. Finally, we present some concluding remarks in Section 7.

\section{Quartic quasi-topological black holes with power-Yang-Mills source} 

\begin{figure}[h]
   	\centering
   	\includegraphics[width=0.8\textwidth]{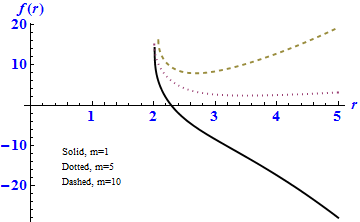}
   	\caption{Dependence of function $f(r)$ (Eq. (\ref{16h})) on the mass for fixed values of $d=7$, $Q=2$, $q=2$, $k=1$, $\mu_2=-0.09$, $\mu_3=-0.006$, $\mu_4=0.0004$ and $\Lambda=-1$.}\label{Asaif1}
   \end{figure}

\begin{figure}[h]
	\centering
	\includegraphics[width=0.8\textwidth]{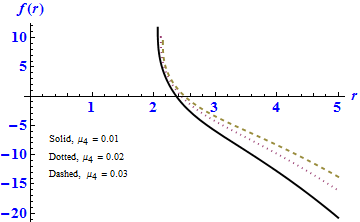}
	\caption{Plot of function $f(r)$ (Eq. (\ref{16h})) for different values of parameter $\mu_4$ and fixed values $d=7$, $m=1$, $Q=2$, $q=2$, $k=1$, $\mu_2=-0.09$, $\mu_3=-0.006$ and $\Lambda=-1$.}\label{Asaif2}
\end{figure}

\begin{figure}[h]
	\centering
	\includegraphics[width=0.8\textwidth]{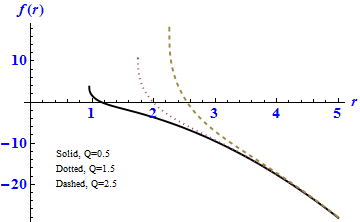}
	\caption{Dependence of function $f(r)$ (Eq. (\ref{16h})) on the Yang-Mills charge $Q$ for fixed values of $d=7$, $m=1$ $q=2$, $k=1$, $\mu_2=-0.09$, $\mu_3=-0.006$, $\mu_4=0.0004$ and $\Lambda=-1$.}\label{Asaif3}
\end{figure}

\begin{figure}[h]
	\centering
	\includegraphics[width=0.8\textwidth]{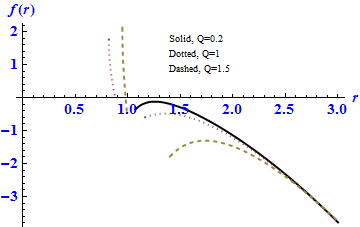}
	\caption{Dependence of function $f(r)$ (Eq. (\ref{16h})) on the Yang-Mills charge $Q$ for fixed values of $q=d_1/4$, $m=1$, $q=2$, $k=1$, $\mu_2=-0.06$, $\mu_3=-0.1$, $\mu_4=0.03$ and $\Lambda=1$.}\label{Asaif4}
\end{figure}

\begin{figure}[h]
	\centering
	\includegraphics[width=0.8\textwidth]{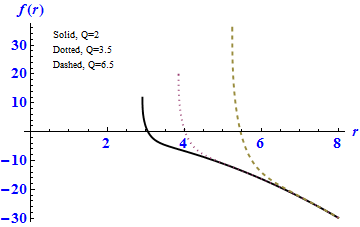}
	\caption{Dependence of function $f(r)$ (Eq. (\ref{16h})) on the Yang-Mills charge $Q$ for fixed values of $d=9$, $q=5$, $m=10$, $k=1$, $\mu_2=-0.06$, $\mu_3=-0.1$, $\mu_4=0.03$ and $\Lambda=0$.}\label{Asaif5}
\end{figure}

The action describing the quartic quasi-topological gravity coupled to Yang-Mills theory is given in Ref. \cite{31W}. Here, we use the power-Yang-Mills field as a source and work for new quartic quasi-topological solutions. In this setup, we consider the N-parameters gauge group $\mathfrak{G}$ whose structure constants $C_{(i)(j)}^{(k)}$ are defined as
\begin{equation}
\gamma_{ij}=-\frac{\Gamma_{(i)(j)}}{|\Gamma|^{1/N}},\label{1h}
\end{equation} 
where, $i.j,k$ run from 1 to $N$, $\Gamma_{(i)(j)}=C^{(k)}_{(i)(l)}C^{(l)}_{(j)(k)}$ and $\Gamma$ is its determinant. Thus, the action for the quartic quasi-topological gravity coupled with the power-Yang-Mills theory in higher spacetime dimensions is given by
 \begin{equation}
\mathcal{I}_{bulk}=\frac{1}{16\pi}\int d^{d}x\sqrt{-g}\bigg[R-2\Lambda+\tilde{\mu}_2\mathfrak{L}_2+\tilde{\mu}_3\mathfrak{L}_3+\tilde{\mu}_4\mathfrak{L}_{4}-\digamma^q\bigg],
\label{2h}
\end{equation}
where $\digamma$ is the Yang-Mills invariant defined as
\begin{equation}
\digamma=\gamma_{ab}F^{(a)}_{\alpha\beta}F^{(b)\alpha\beta}.\label{3h}
\end{equation}
 Also, $\Lambda$ denotes the cosmological constant, $R$ refers to the Ricci scalar and $\tilde{\mu}_2$, $\tilde{\mu}_3$ and $\tilde{\mu}_4$ are the coefficients of quasi-topological gravity. Furthermore, $\mathfrak{L}_2$, $\mathfrak{L}_3$ and $\mathfrak{L}_4$ denote the Lagrangians for Gauss-Bonnet, the cubic and quartic quasi-topological theories, respectively, and are expressed as \cite{31W}
\begin{equation}
\mathfrak{L}_2=R_{\mu\nu\gamma\rho}R^{\mu\nu\gamma\rho}-4R_{\mu\nu}R^{\mu\nu}+R^2,
\label{4h}
\end{equation}
\begin{equation}\begin{split}
\mathfrak{L}_3&=R^{\rho\sigma}_{\mu\nu}R^{\alpha\beta}_{\rho\sigma}R^{\mu\nu}_{\alpha\beta}+\frac{1}{8(2d-3)(d-4)}\big(b_1R_{\mu\nu\rho\sigma}R^{\mu\nu\rho\sigma}R+b_2R_{\mu\nu\rho\sigma}R^{\mu\nu\rho}_{\alpha}R^{\sigma\alpha}\\&+b_3R_{\mu\nu\rho\sigma}R^{\mu\rho}R^{\nu\sigma}+b_4R^{\nu}_{\mu}R^{\rho}_{\nu}R^{\mu}_{\rho}+b_5R_{\mu}^{\nu}R_{\nu}^{\mu}R+b_6R^3\big),
\label{5h}\end{split}
\end{equation}
\begin{equation}\begin{split}
\mathfrak{L}_4&=c_1R_{\mu\nu\rho\sigma}R^{\rho\sigma\alpha\beta}R^{\kappa\gamma}_{\alpha\beta}R_{\kappa\gamma}^{\mu\nu}+c_2R_{\mu\nu\rho\sigma}R^{\mu\nu\rho\sigma}R^{\alpha\beta}_{\alpha\beta}+c_3RR_{\mu\nu}R^{\mu\rho}R_{\rho}^{\nu}\\&+c_4\big(R_{\mu\nu\rho\sigma}R^{\mu\nu\rho\sigma}\big)^2+c_5R_{\mu\nu}R^{\mu\rho}R_{\rho\sigma}R^{\sigma\nu}+c_6RR_{\mu\nu\rho\sigma}R^{\mu\rho}R^{\nu\sigma}\\&+c_7R_{\mu\nu\rho\sigma}R^{\mu\rho}R^{\nu\alpha}R^{\sigma}_{\alpha}+c_8R_{\mu\nu\rho\sigma}R^{\mu\rho\alpha\beta}R^{\nu}_{\alpha}R^{\sigma}_{\beta}+c_9R_{\mu\nu\rho\sigma}R^{\mu\rho}R_{\alpha\beta}R^{\nu\alpha\sigma\beta}\\&+c_{10}R^4+c_{11}R^2R_{\mu\nu\rho\sigma}R^{\mu\nu\rho\sigma}+c_{12}R^2R_{\alpha\beta}R^{\alpha\beta}+c_{13}R_{\mu\nu\rho\sigma}R^{\mu\nu\alpha\beta}R_{\alpha\beta\gamma}^{\nu}R^{\sigma\gamma}\\&+c_{14}R_{\mu\nu\rho\sigma}R^{\mu\alpha\rho\beta}R_{\kappa\alpha\gamma\beta}R^{\kappa\nu\gamma\sigma},
\label{6h}\end{split}
\end{equation}
where the coefficients $b_i$'s and $c_i$'s are given in the Appendix. The Yang-Mills gauge field can be defined as
\begin{equation}
F^{(a)}=dA^{(a)}+\frac{1}{2 \overline{\eta}}C^{(a)}_{(b)(c)}A^{(b)}\wedge A^{(c)},\label{7h}
\end{equation}  
where $\overline{\eta}$ is the coupling constant while $A^{(a)}$ refers to Yang-Mills potential of the $SO(d-1)$ gauge group. The structure constants have been computed in Ref. \cite{55W}. We take the metric ansatz in $d$-dimensional spacetime as
\begin{equation}
ds^2=-f(r)dt^2+\frac{dr^2}{f(r)}+r^2d\Omega^2_k,\label{8h}
\end{equation}
where
\begin{equation}\begin{split}
d\Omega^2_k=\left\{ \begin{array}{rcl}
d\theta_1^2+\sum_{j=2}^{d-2}\prod_{l=1}^{j-1}\sin^2\theta_l d\theta_j^2, & 
& k=1, \\d\theta_1^2+\sinh^2\theta_1d\theta_2^2+\sinh^2\theta_1\sum_{j=3}^{d-2}\prod_{l=2}^{j-1}\sin^2\theta_ld\theta^2_j, &  & k=-1,\\\sum_{j=1}^{d-2}d\phi_j^2, & & k=0,
\end{array}\right.\label{9h}
\end{split}
\end{equation}
stands for the metric of a $(d-2)$-dimensional hyper-surface of constant curvature $(d-2)(d-3)k$ and volume $\mathcal{V}_{d-2}$.
 Now, using the line element (\ref{8h}) and the Lagrangian density of power-Yang-Mills model i.e. $\mathfrak{L}_{pYM}=-\digamma^q$, it is possible to write the energy-momentum tensor associated with power-Yang-Mills field as
\begin{equation}
T^{(a)\nu}_{\mu}=-\frac{1}{2}\bigg[\delta^{\nu}_{\mu}\digamma^q-4q \sum_{a=1}^{(d-2)(d-1)/2}\bigg(F^{(a)}_{\mu\lambda}F^{(a)\nu\lambda}\bigg)\digamma^{q-1}\bigg].\label{10h}
\end{equation}
The variation of action (\ref{2h}) with respect to the gauge potentials $A^{(a)}$ yields
\begin{equation} 
d(^{\star}F^{(a)}\digamma^{q-1})+\frac{1}{\eta}C^{(a)}_{(b)(c)}\digamma^{q-1}A^{(b)}\wedge^{\star}F^{(c)}=0,  \label{11h}
\end{equation}
where $\star$ denotes the duality. Now, using the line element (\ref{8h}) and the Wu-Yang ansatz introduced in Refs. \cite{30W,31W,32W,33W}, the power-Yang-Mills field equations \cite{56W,57W} will be satisfied provided the gauge potential one-forms are expressed as 
\begin{equation}
A^{(a)}=\frac{Q}{r^2}C^{(a)}_{(l)(j)}x^{l}dx^{j}, r^2=\sum_{l=1}^{d-1}x_{l}^2.   \label{12h}
\end{equation}
The parameter $Q$ is proportional to the Yang-Mills magnetic charge while $2\leq j+1\leq l\leq d-1$. For simplicity it is convenient to use $d_i=d-i$ and redefine the quasi-topological coefficients as
\begin{align}\begin{split}
&\mu_2=d_3d_4\tilde{\mu}_2,\\&
\mu_3=\frac{d_3d_6(3d_1^2-9d_1+4)}{8(2d-3)}\tilde{\mu}_3,\\&\mu_4=d_1d_2d_4d_8d_3^2(d_1^4-15d_1^4+17d_1^3-156d_1^2+150d_1-42)\tilde{\mu}_4.\label{13h}\end{split}
\end{align} 
The equations of motion describing gravitational field can be obtained if we vary the action (\ref{2h}) with respect to metric tensor $g_{\mu\nu}$. Thus, using Eqs. (\ref{11h}), (\ref{12h}) and (\ref{13h}) in (\ref{2h}) one can get the following fourth-order equation 
\begin{equation}
\mu_4\Phi^4+\mu_3\Phi^3+\mu_2\Phi^2+\Phi+\Upsilon=0,\label{14h}
\end{equation}
where $\Phi=(k-f(r))/r^2$ and
\begin{equation}\begin{split}
\Upsilon=\left\{ \begin{array}{rcl}
-\frac{2\Lambda}{d_1d_2}-\frac{m}{r^{d_1}}-\frac{d_3^qd_2^{q-1}Q^{2q}}{(d_1-4q)r^{4q}}, & 
& q\neq\frac{d_1}{4} \\-\frac{2\Lambda}{d_1d_2}-\frac{m}{r^{d_1}}-\frac{Q^{d_1/2}d_3}{r^{d_1}}\ln{r}, &  & q=\frac{d_1}{4}.
\end{array}\right.\label{15h}
\end{split}
\end{equation}
The constant of integration $m$ in the above equation refers to the mass of gravitating object with power-Yang-Mills magnetic charge. The power-Yang-Mills quasi-topological solution can be explicitly expressed from the polynomial equation (\ref{14h}) as
\begin{equation}\begin{split}
f(r)=k-r^2\times\left\{ \begin{array}{rcl}
-\frac{\mu_3}{4\mu_4}-\frac{1}{2}\big(H_1-\sqrt{\frac{2A_2}{H_1}-2\overline{y}-3A_1}\big), & 
& \mu_4>0 \\-\frac{\mu_3}{4\mu_4}+\frac{1}{2}\big(H_1-\sqrt{-\frac{2A_2}{H_1}-2\overline{y}-3A_1}\big), &  & \mu_4<0,
\end{array}\right.\label{16h}
\end{split}
\end{equation}
where
\begin{align}\begin{split}
&H_1=\big(A_1+2\overline{y}\big)^{\frac{1}{2}}, \\& A_1=\frac{\mu_2}{4}-\frac{3\mu_3^2}{8\mu_4^2},\\&
H_2=-\frac{A_1^3}{108}+\frac{A_1\mathcal{Z}}{3}-\frac{A_2^2}{8},\\& A_2=\frac{\mu_3^3}{8\mu_4^3}-\frac{\mu_3\mu_2}{2\mu_4^2}+\frac{1}{\mu_4},\\&\mathcal{Z}=-\frac{3\mu_3^4}{256\mu_4^4}+\frac{\mu_2\mu_3^2}{16\mu_4^3}-\frac{\mu_3}{4\mu_4^2}+\frac{\Upsilon}{\mu_4},\\& \mathcal{W}=\bigg(-\frac{H_2}{2}\pm\sqrt{\frac{H_2^2}{4}+\frac{\overline{P}^3}{27}}\bigg)^{\frac{1}{3}}, \\& \overline{P}=-\frac{A_1^2}{12}-\mathcal{Z},\label{17h}\end{split}
\end{align}
and 
\begin{equation}\begin{split}
\overline{y}=\left\{ \begin{array}{rcl}
-\frac{5}{6}A_1+\mathcal{W}-\frac{\overline{P}}{3\mathcal{W}}, & 
& W\neq 0 \\-\frac{5}{6}A_1+\mathcal{W}-H_2^{\frac{1}{3}}, &  & \mathcal{W}=0.
\end{array}\right.\label{18h}
\end{split}
\end{equation}
It can be easily understood from (\ref{16h}) that the metric function describes solutions of the gravitational field equations of two types for $\mu_4>0$ and $\mu_4<0$. Similar to the case of Yang-Mills black holes \cite{31W} in this theory, in this situation too, the parameter $\Upsilon$ obtained in (\ref{15h}) becomes highly negative for small values of the coordinate $r$. This makes the fourth term in parameter $\mathcal{Z}$ as well as the  parameter $\overline{P}$ of (\ref{17h}), very large. In case $\mu_4<0$, this gives negatively large value for $\overline{P}$ which yields an imaginary solution for $\mathcal{W}$ for small values of $r$. Hence, we would not consider $\mu_4<0$. In order to get the metric function (\ref{16h}) in simpler form, we assume the special case $\overline{\mu}_2=\overline{\mu}_3=0$. Thus, the power-Yang-Mills quasi-topological solution for $\mu_4\neq 0$ becomes
\begin{equation}
f(r)=k-\frac{r^2}{2}\bigg[\mp\sqrt{2\Delta^{\frac{1}{3}}+\frac{2\Upsilon}{3\mu_4\Delta^{\frac{1}{3}}}}\pm\sqrt{-2\Delta^{\frac{1}{3}}\pm\frac{2}{\mu_4}\bigg(2\Delta^{\frac{1}{3}}+\frac{2\Upsilon}{3\mu_4\Delta^{\frac{1}{3}}}\bigg)^{-\frac{1}{2}}-\frac{2\Upsilon}{3\mu_4\Delta^{\frac{1}{3}}}}\bigg],\label{19h}
\end{equation} 
where
\begin{equation}
\Delta=\frac{1}{16\mu_4^2}+\sqrt{\frac{1}{256\mu_4^4}-\frac{\Upsilon^3}{27\mu_4^3}}.\label{20h}
\end{equation}
It should be noted that the upper sign in Eq. (\ref{19h}) corresponds to the case $\mu_4>0$ while the lower sign is for $\mu_4<0$. In the limit $\mu_4\rightarrow 0$, we can write the series expansion of metric function (\ref{19h}) as
\begin{equation}
f(r)=k+\Upsilon r^2+\mu_4\Upsilon^4r^2+4\mu_4^2\Upsilon^7r^2+O((\mu_4)^{8/3}),\label{21h}
\end{equation}
where $\Upsilon$ is given in Eq. (\ref{15h}). The above expansion implies the Einstein-power-Yang-Mills solutions with some corrections in $\mu_4$. The Ricci and Kretschmann invariants for the metric ansatz (\ref{8h}) respectively take the forms
\begin{eqnarray}\begin{split}
R&=\bigg[d_2d_3\bigg(\frac{k-f(r)}{r^2}\bigg)-f''(r)-\frac{2d_2}{r}f'(r)\bigg],\label{22h}\end{split}
\end{eqnarray}
and
\begin{eqnarray}\begin{split}
K&=\bigg[2d_2d_3\bigg(\frac{k-f(r)}{r^2}\bigg)^2-\big(f''(r)\big)^2+\frac{2d_2}{r^2}\big(f'(r)\big)^2\bigg].\label{23h}\end{split}
\end{eqnarray}
The primes denote derivatives with respect to the coordinate $r$. So, by using the metric function obtained for the case $\mu_4>0$, it can easily be shown that both the scalars diverge at the center $r=0$. Hence, there is a true curvature singularity at $r=0$ for our power-Yang-Mills solutions. The horizons of the black hole can be described from the condition $f(r_+)=0$. Fig. \ref{Asaif1} shows the plot of the metric function for different values of the mass parameter $m$. The values of $r$ for which the curve touches the horizontal axis correspond to the horizon's location. It should be noted that the metric function (\ref{16h}) can be interpreted as power-Yang-Mills quasi-topological black hole with two horizons when $m>m_{ext}$; extremal black hole when $m=m_{ext}$ and naked singularities otherwise. Fig. \ref{Asaif2} describes the behaviour of solution (\ref{16h}) for various values of $\mu_4$. It can be observed that for the fixed values of parameters $d$, $m$, $\Lambda$, $k$, $Q$, and $q$, the values of horizons are affected by the parameter $\mu_4$. However, at infinity the behaviour of the metric function does not depend on parameter $\mu_4$. Furthermore, the dependence of the solution on the Yang-Mills charge $Q$ is presented in Fig. \ref{Asaif3}. It is easily seen that for the chosen fixed values of other parameters, the outer horizon is independent of the charge parameter $Q$ whereas the inner one increases with the increases of $Q$. Similarly, the behaviour of the resulting metric function associated with dS black holes for different values of Yang-Mills charge $Q$ in spacetime dimensions such that $q=d_1/4$ is shown in Fig. \ref{Asaif4}.  Also, the behaviour of the metric function corresponding to asymptotically flat black holes i.e. with $\Lambda=0$ for different values of $Q$ is shown in Fig. \ref{Asaif5}. It is also worthwhile to note that, for $q=1$, the solution (\ref{16h}) reduces to the metric function of Yang-Mills quasi-topological black hole \cite{31W}.

\section{Thermodynamics of quartic quasi-topological power-Yang-Mills black holes} 
 
 Now, we study the thermodynamic properties of the black holes described by equations (\ref{14h})-(\ref{18h}). We can compute the Arnowitt Deser Misner mass density with the help of subtraction method \cite{58W} as follows
 \begin{equation}
 M=\frac{d_2}{16\pi}m,\label{24h}
 \end{equation}
 where the parameter $m$ is given in Eq. (\ref{15h}). Hence, we can write the value of $M$ in terms of the outer horizon radius as 
  \begin{equation}\begin{split}
 M&=\left\{ \begin{array}{rcl}
 \frac{d_2}{16\pi}\bigg(\mu_4 k^4r_+^{d_9}+\mu_3k^3r_+^{d_7}+\mu_2k^2r_+^{d_5}+kr_+^{d_3}-\frac{2\Lambda r_+^{d_1}}{d_1d_2}-\frac{d_2^{q-1}d_3^qQ^{2q}r_+^{d_1-4q}}{(d_1-4q)}\bigg), & 
 & q\neq\frac{d_1}{4}, \\\frac{d_2}{16\pi}\bigg(\mu_4 k^4r_+^{d_9}+\mu_3k^3r_+^{d_7}+\mu_2k^2r_+^{d_5}+kr_+^{d_3}-\frac{2\Lambda r_+^{d_1}}{d_1d_2}-Q^{d_1/2}d_3\ln{r_+}\bigg), &  & q=\frac{d_1}{4}.
 \end{array}\right.\label{25h}\end{split}
  \end{equation} 
 The Yang-Mills charge corresponding to this black hole can be computed through the Gauss law as
 \begin{equation}
 \tilde{Q}=\frac{1}{4\pi\sqrt{d_2d_3}}\int d^{d-2}r\sqrt{\sum_{a=1}^{d_2d_1/2}\bigg(F^{(a)}_{\mu\lambda}F^{(a)\nu\lambda}\bigg)}=\frac{Q}{4\pi}.\label{26h}
 \end{equation}
 Using the condition $f(r_+)=0$ and differentiating the polynomial equation (\ref{14h}) yields the Hawking temperature as
 \begin{equation}\begin{split}
 T_{H}(r_+)&=\frac{f'(r_+)}{4\pi}=\left\{ \begin{array}{rcl}
 \frac{r_+^8}{\mathcal{W}_2}\bigg(\frac{d_1\mathcal{W}_1}{r_+^9}-\frac{2\Lambda }{d_2r_+}-\frac{d_2^{q-1}d_3^qQ^{2q}}{r_+^{4q+1}}\bigg)-\frac{2k}{r_+}, & 
 & q\neq\frac{d_1}{4}, \\\frac{r_+^8}{\mathcal{W}_2}\bigg(\frac{d_1\mathcal{W}_1}{r_+^9}-\frac{2\Lambda }{d_2r_+}-\frac{d_3Q^{d_1/2}}{r_+^{d}}\bigg)-\frac{2k}{r_+}, &  & q=\frac{d_1}{4},
 \end{array}\right.\label{27h}\end{split}
 \end{equation} 
 where
 \begin{align}\begin{split}
 &\mathcal{W}_1=\mu_4k^4+\mu_3k^3r_+^2+\mu_2k^2r_+^4+kr_+^6,\\&
 \mathcal{W}_2=4\mu_4k^3+3\mu_3k^2r_+^2+2k\mu_2r_+^4+r_+^6.\label{28h}\end{split}
 \end{align} 
  \begin{figure}[h]
 	\centering
 	\includegraphics[width=0.8\textwidth]{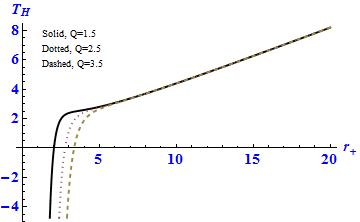}
 	\caption{Plot of temperature $T_H(r)$ (Eq. (\ref{27h})) for different values of the charge parameter $Q$ and fixed values of $d=7$, $q=2$, $k=1$, $\mu_2=-0.06$, $\mu_3=-0.1$, $\mu_4=0.03$ and $\Lambda=-1$.}\label{Asaif6}
 \end{figure}
 \begin{figure}[h]
 	\centering
 	\includegraphics[width=0.8\textwidth]{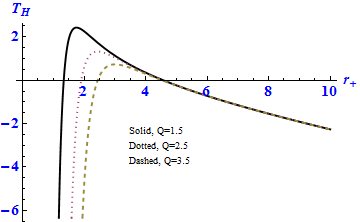}
 	\caption{Dependence of temperature $T_H(r)$ (Eq. (\ref{27h})) on the charge parameter $Q$ for fixed values of $d=9$, $q=2$, $k=1$, $\mu_2=-0.06$, $\mu_3=-0.1$, $\mu_4=0.03$ and $\Lambda=1$.}\label{Asaif7}
 \end{figure}
 Fig. \ref{Asaif6} describes the behaviour of temperature for several values of charge $Q$ when $\Lambda=-1$. The positivity of temperature indicates that the black hole solution is physical. Similarly, the behaviour of Hawking temperature for the case $q=d_1/4$ and $\Lambda=1$ can be observed from Fig. \ref{Asaif7}. It can be easily seen from these graphs that the solutions with negative $\Lambda$ may have a larger range of parameters with positive temperature than the ones with positive $\Lambda$. Following Ref. \cite{59W}, we compute the entropy density as
 \begin{equation}
 \mathcal{S}=\frac{r_+^{d_2}}{4}+\frac{d_2k\mu_2r_+^{d_4}}{2d_4}+\frac{3d_2k^2\mu_3r_+^{d_6}}{4d_6}+\frac{d_2k^3\mu_4r_+^{d_8}}{4d_8}.\label{29h}
 \end{equation}
 Consideration of mass $M$ as a function of entropy $\mathcal{S}$ and charge $\tilde{Q}$ enables us to construct the first law as
 \begin{equation}
 dM=T_Hd\mathcal{S}+\mathcal{U}d\tilde{Q},\label{30h}
 \end{equation}
 where $T_H=\bigg(\frac{\partial M}{\partial \mathcal{S}}\bigg)_{\tilde{Q}}$ and $\mathcal{U}=\bigg(\frac{\partial M}{\partial \tilde{Q}}\bigg)_{\mathcal{S}}$. The calculations show that the same form of temperature i.e. (\ref{27h}) can be obtained from this relation. Moreover, the power-Yang-Mills potential can be obtained as follows:
 \begin{equation}\begin{split}
 \mathcal{U}&=\bigg(\frac{\partial M}{\partial \tilde{Q}}\bigg)_{\mathcal{S}}=\left\{ \begin{array}{rcl}
 -\frac{qd_2^qd_3^q(4\pi \tilde{Q})^{2q-1}}{8\pi(d_1-4q)}r_+^{d_1-4q}, & 
 & q\neq\frac{d_1}{4}, \\-\frac{d_1d_2d_3(4\pi\tilde{Q})^{d_3/2}}{32\pi}\ln{r_+}, &  & q=\frac{d_1}{4},
 \end{array}\right.\label{31h}\end{split}
 \end{equation} 
The charge $\tilde{Q}$ should be fixed in the canonical ensemble and the thermodynamic stability can be examined by assuming small variations of the entropy. Hence, thermodynamic stability would be guaranteed if the specific heat is positive. The specific heat capacity at a constant Yang-Mills charge $\tilde{Q}$ can be computed from $C_H=T_H(\frac{\partial \mathcal{S}}{\partial T_H})_{\tilde{Q}}$. So, from Eqs. (\ref{27h})-(\ref{29h}), we can calculate the heat capacity in the form
 \begin{equation} \begin{split}
 C_H=\left\{ \begin{array}{rcl}
 \frac{\mathcal{W}_2\mathcal{W}_3\big(d_1d_2\mathcal{W}_1r_+^{4q}-2\Lambda r_+^{4q+8}-(4\pi\tilde{Q})^{2q}d_2^qd_3^qr_+^8-2kd_2\mathcal{W}_2r_+^{4q}\big)}{4d_2r_+^7\big(r_+^{4q-6}\mathcal{W}_2^2G(r_+)+((4q-7)\mathcal{W}_2+r\mathcal{W}_2'(r_+))(4\pi\tilde{Q})^{2q}d_2^{q-1}d_3^q)\big)}, & 
 & q\neq\frac{d_1}{4}, \\\frac{\mathcal{W}_2\mathcal{W}_3\big(d_1d_2\mathcal{W}_1r_+^{d}-2\Lambda r_+^{d+8}-d_2d_3(4\pi\tilde{Q})^{d_1/2}r_+^9-2k\mathcal{W}_2d_2r_+^d\big)}{4d_2r_+^8\big(r_+^{d_7}\mathcal{W}_2^2G(r_+)+(r_+\mathcal{W}_2'+d_8\mathcal{W}_2)d_3(4\pi\tilde{Q})^{d_1/2}\big)}, &  & q=\frac{d_1}{4},
 \end{array}\right. \label{32h}
 \end{split}\end{equation}
 where 
 \begin{align}\begin{split}
 &\mathcal{W}_1'(r_+)=2\mu_3k^3r_++4\mu_2k^2r_+^3+6kr_+^5,\\&
 \mathcal{W}_2'(r_+)=6\mu_3k^2r_++8k\mu_2r_+^3+6r_+^5,\\&\mathcal{W}_3(r_+)=d_2r_+^{d_3}+2d_2k\mu_2r_+^{d_5}+3d_2k^2\mu_3r_+^{d_7}+d_2k^3\mu_4r_+^{d_9}\label{33h}\end{split}
 \end{align} 
 and
 \begin{equation} \begin{split}
  G(r_+)&=\frac{1}{\mathcal{W}_2}\bigg(\frac{d_1\mathcal{W}_1'}{r_+}-\frac{d_1\mathcal{W}_1}{r_+^2}-\frac{14\Lambda r_+^6}{d_2}\bigg)+\frac{2k}{r_+^2}-\frac{\mathcal{W}_2'}{\mathcal{W}_2^2}\bigg(\frac{d_1\mathcal{W}_1}{r_+}-\frac{2\Lambda r_+^7}{d_2}\bigg).\label{34h}
 \end{split}\end{equation}
 \begin{figure}[h]
 	\centering
 	\includegraphics[width=0.8\textwidth]{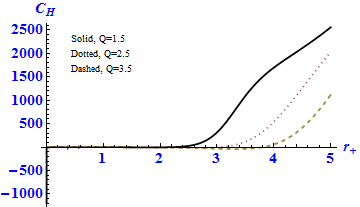}
 	\caption{Dependence of heat capacity $C_H$ (Eq. (\ref{32h})) on the Yang-Mills charge $\tilde{Q}=Q/4\pi$ for fixed values of $d=7$, $q=2$, $k=1$, $\mu_2=-0.06$, $\mu_3=-0.1$, $\mu_4=0.03$ and $\Lambda=-1$.}\label{Asaif8}
 \end{figure}
\begin{figure}[h]
	\centering
	\includegraphics[width=0.8\textwidth]{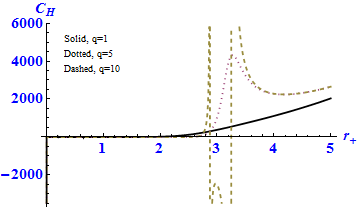}
	\caption{Dependence of heat capacity $C_H$ (Eq. (\ref{32h})) on the parameter $q$ for fixed values of $d=7$, $Q=1.5$, $k=1$, $\mu_2=-0.06$, $\mu_3=-0.1$, $\mu_4=0.03$ and $\Lambda=-1$.}\label{Asaif9}
\end{figure}
\begin{figure}[h]
	\centering
	\includegraphics[width=0.8\textwidth]{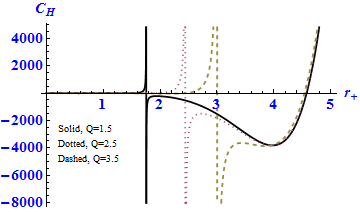}
	\caption{Dependence of heat capacity $C_H$ (Eq. (\ref{32h})) on the Yang-Mills charge $\tilde{Q}=Q/4\pi$ for fixed values of $d=9$, $q=d_1/4=2$, $k=1$, $\mu_2=-0.06$, $\mu_3=-0.1$, $\mu_4=0.03$ and $\Lambda=1$.}\label{Asaif10}
\end{figure}
The behaviour of heat capacity depending on outer horizon $r_+$ for various values of charge $\tilde{Q}=Q/4\pi$ is given in Fig. \ref{Asaif8}. The region where this quantity is positive implies black hole stability in this ensemble. It can also be observed that as charge $Q$ increases, the outer horizon for stable black hole increases. Similarly, Fig. \ref{Asaif9} shows the corresponding plot for different values of parameter $q$. The case $q=1$ corresponds to the heat capacity of Yang-Mills black hole in this gravity theory. It can be observed that this parameter of nonlinear Yang-Mills field affects the thermodynamic local stability of black holes. The plot of heat capacity for different values of charge $Q$ with positive cosmological constant and spacetime dimensions satisfying $q=d_1/4$ is shown in Fig. \ref{Asaif10}. The region of local stability and instability can easily be seen from it. It is also worthwhile to note that the points at which this quantity vanishes indicate the possibility of first order phase transitions. However, those values for which it is infinite correspond to the possibility of second order phase transitions. When it comes to the grand canonical ensemble, both the charge $Q$ and entropy $\mathcal{S}$ should be treated as variables. In addition to specific heat and Hawking temperature, the local thermodynamic stability can be guaranteed from the positivity of $\partial^2M/\partial\tilde{Q}^2$ and the determinant of the Hessian matrix \cite{60W,61W}. The determinant of the Hessian matrix is given by
     \begin{equation} \begin{split}
    det\textbf{H}&=\bigg(\frac{\partial^2M}{\partial\mathcal{S}^2}\bigg)\bigg(\frac{\partial^2M}{\partial\tilde{Q}^2}\bigg)-\bigg(\frac{\partial^2M}{\partial\mathcal{S}\partial\tilde{Q}}\bigg)^2.\label{35h}
    \end{split}\end{equation} 
  If we compute $\bigg(\frac{\partial^2M}{\partial\tilde{Q}^2}\bigg)$ for our resulting solution, we have
    \begin{equation}\begin{split}
   \bigg(\frac{\partial^2M}{\partial\tilde{Q}^2}\bigg)&=\left\{ \begin{array}{rcl}
    -\frac{q(2q-1)d_2^qd_3^q(4\pi)^{2q} (\tilde{Q})^{2q-2}}{8\pi(d_1-4q)}r_+^{d_1-4q}, & 
    & q\neq\frac{d_1}{4}, \\-\frac{d_1d_2d_3^2(4\pi)^{d_3/2}(\tilde{Q})^{d_5/2}}{64\pi}\ln{r_+}, &  & q=\frac{d_1}{4}.
    \end{array}\right.\label{36h}\end{split}
    \end{equation}
   This equation shows that the parameter $\bigg(\frac{\partial^2M}{\partial\tilde{Q}^2}\bigg)$ is negative for $d_1> 4q$ when $2q\geq 1$ and so the black hole would be unstable in this ensemble. However, for the spacetime dimensions satisfying $d_1<4q$ it is positive and so thermal stability will be determined from the behaviour of the Hessian matrix. It should be noted that, when the spacetime dimensions satisfy $q=d_1/4$, then the black hole is unstable in grand canonical ensemble. Furthermore, the case $q=1$ in Eq. (\ref{36h}) corresponds to the instability of Yang-Mills quasi-topological black hole \cite{31W}. One can compute the Hessian matrix determinant as 
   \begin{equation}\begin{split}
   det\textbf{H}&=\left\{ \begin{array}{rcl}
  \frac{(4\pi)^{2q}(2q)(2q-1)d_2^qd_3^q\tilde{Q}^{2q-2}r_+^{d_1-4q}A_q(r_+)}{\big(d_2r_+^{d_3}+2d_2k\mu_2r_+^{d_5}+3d_2\mu_3k^2r_+^{d_7}+d_2k^3\mu_4r_+^{d_9}\big)}-\frac{4q^2(4\pi)^{2q}d_2^{2q-2}d_3^{2q}\tilde{Q}^{4q-2}}{\mathcal{W}_2^2r_+^{8q-14}}, & 
   & q>\frac{d_1}{4}, \\\frac{d_1d_2d_3(4\pi)^{d_1/2}\tilde{Q}^{d_3/2}\ln{r_+}A_q(r_+)}{32\pi\big(d_2r_+^{d_3}+2d_2k\mu_2r_+^{d_5}+3d_2\mu_3k^2r_+^{d_7}+d_2k^3\mu_4r_+^{d_9}\big)}-\frac{d_1^2d_3^2(4\pi)^{d_1}\tilde{Q}^{d_3}}{4r_+^{2d_8}\mathcal{W}_2^2}, &  & q=\frac{d_1}{4},
   \end{array}\right.\label{36AH}\end{split}
   \end{equation}
   where
    \begin{equation}\begin{split}
   A_q(r_+)&=\left\{ \begin{array}{rcl}
   \frac{\mathcal{W}_2^2G(r_+)r_+^{4q-6}+(4\pi)^{2q}d_2^{q-1}d_3^q\tilde{Q}^{2q}\big((4q-7)\mathcal{W}_2(r_+)+r_+\mathcal{W}_2'(r_+)\big)}{4\pi \mathcal{W}_2^2(4q-d_1)r_+^{4q-6}}, & 
   & q>\frac{d_1}{4}, \\-\frac{\mathcal{W}_2^2G(r_+)r_+^{d_7}+(4\pi)^{d_1/2}d_3\tilde{Q}^{d_1/2}\big(r_+\mathcal{W}_2'(r_+)+)+d_8\mathcal{W}_2(r_+)\big)}{8\pi r_+^{d_7}\mathcal{W}_2^2(r_+)}, &  & q=\frac{d_1}{4}.
   \end{array}\right.\label{36AH1}\end{split}
   \end{equation}
   \begin{figure}[h]
   	\centering
   	\includegraphics[width=0.8\textwidth]{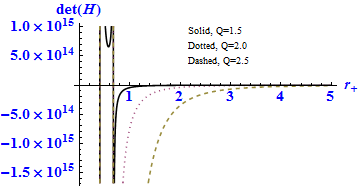}
   	\caption{Plot of $det\textbf{H}$ (Eq. (\ref{36AH})) for different values of the Yang-Mills charge $\tilde{Q}=Q/4\pi$ and fixed values of $d=11$, $q=3$, $k=1$, $\mu_2=-0.06$, $\mu_3=-0.1$, $\mu_4=0.03$ and $\Lambda=-1$.}\label{AAli1}
   \end{figure}
  \begin{figure}[h]
	\centering
	\includegraphics[width=0.8\textwidth]{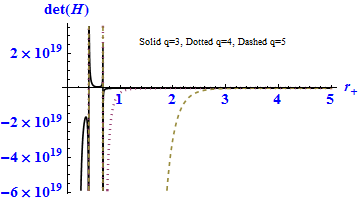}
	\caption{Dependence of $det\textbf{H}$ (Eq. (\ref{36AH})) on the parameter $q$ for fixed values of $d=11$, $Q=3$, $k=1$, $\mu_2=-0.06$, $\mu_3=-0.1$, $\mu_4=0.03$ and $\Lambda=-1$.}\label{AAli2}
\end{figure}
  \begin{figure}[h]
	\centering
	\includegraphics[width=0.8\textwidth]{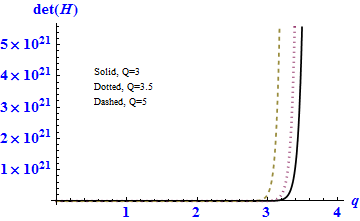}
	\caption{Plot of $det\textbf{H}$ (Eq. (\ref{36AH})) as a function of the nonlinearity parameter $q$ for fixed values of $d=11$, $r_+=0.5$, $k=1$, $\mu_2=-0.06$, $\mu_3=-0.1$, $\mu_4=0.03$ and $\Lambda=-1$.}\label{AAli3}
\end{figure}
Figs. \ref{AAli1}-\ref{AAli2} describe the plots of $det\textbf{H}$ in terms of the outer horizon when $q>d_1/4$. These plots indicate that the power-Yang-Mills black holes of smaller outer horizons can be thermodynamically stable in the grand canonical ensemble because the associated determinant of the Hessian matrix could be positive when $q>d_1/4$. However, as $r_+$ increases this quantity is negative and so we have instability of black holes. Fig. \ref{AAli3} shows the plot of this quantity as a function of the nonlinearity parameter $q$. It may be noted that the parameter $q$ has a great influence on the stability of smaller black holes. One can also confirm from this plot that the Yang-Mills black holes \cite{31W} (i.e. the case corresponding to $q=1$ in power-Yang-Mills black hole solution (\ref{16h})) are thermally unstable in this ensemble.  
  \section{Pure quasi-topological black holes with power-Yang-Mills source} 
 
\begin{figure}[h]
  	\centering
  	\includegraphics[width=0.8\textwidth]{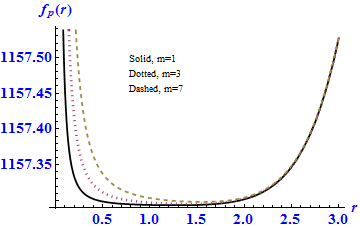}
  	\caption{Dependence of $f_p$ (Eq. (\ref{39h})) on the mass when $d=11$, $q=2$, $k=1$, $Q=5$, $\mu_4=10^{-7}$ and $\Lambda=1$.}\label{Asaif11}
  \end{figure}

\begin{figure}[h]
	\centering
	\includegraphics[width=0.8\textwidth]{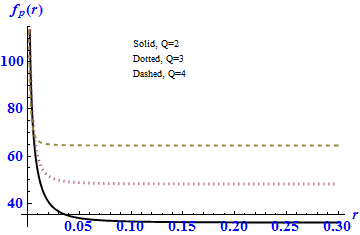}
	\caption{Dependence of $f_p$ (Eq. (\ref{39h})) on the Yang-Mills charge parameter $Q$ when $d=11$, $q=2$, $k=-1$, $m=3$, $\mu_4=0.004$ and $\Lambda=1$.}\label{Asaif12}
\end{figure}

\begin{figure}[h]
	\centering
	\includegraphics[width=0.8\textwidth]{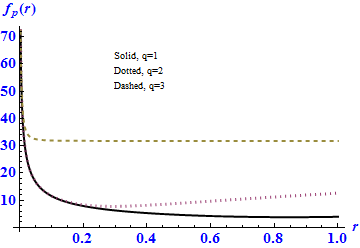}
	\caption{Dependence of $f_p$ (Eq. (\ref{39h})) on the parameter $q$ when $d=11$, $Q=2$, $k=-1$, $m=1$, $\mu_4=0.004$ and $\Lambda=1$.}\label{Asaif13}
\end{figure}

\begin{figure}[h]
	\centering
	\includegraphics[width=0.8\textwidth]{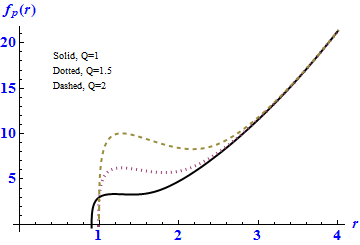}
	\caption{Dependence of $f_p$ (Eq. (\ref{39h})) on the Yang-Mills charge parameter $Q$ when $d=13$, $q=d_1/4=3$, $k=-1$, $m=1$, $\mu_4=0.004$ and $\Lambda=1$.}\label{Asaif14}
\end{figure}

  Now, we want to determine a new class of power-Yang-Mills black hole solutions in pure quasi-topological gravity. In order to do this, we set $R=\mathfrak{L}_2=\mathfrak{L}_3=0$, so that the action becomes
  \begin{equation}
  \mathcal{I}_{bulk}=\frac{1}{16\pi}\int d^{d}x\sqrt{-g}\bigg[-2\Lambda+\tilde{\mu}_4\mathfrak{L}_{4}-\digamma^q\bigg],
  \label{37h}
  \end{equation}
  while, the field equation (\ref{14h}) reduces to 
  \begin{equation}
  \mu_4\Phi^4+\Upsilon=0,\label{38h}
  \end{equation}
  where $\Phi=(k-f(r))/r^2$ and $\Upsilon$ was defined in (\ref{15h}). Hence, the solution in this case can be obtained as
  \begin{equation}\begin{split}
  f_p(r)&=\left\{ \begin{array}{rcl}
  k\mp\frac{r^{\frac{3}{2}}}{\mu_4}\big[\mu_4^3\big(\frac{2\Lambda r^2}{d_1d_2}+\frac{m}{r^{d_3}}+\frac{d_3^qd_2^{q-1}Q^{2q}}{(d_1-4q)r^{4q-2}}\big)\big]^{\frac{1}{4}}, & 
  & q\neq\frac{d_1}{4}, \\k\mp\frac{r^{\frac{3}{2}}}{\mu_4}\big[\mu_4^3\big(\frac{2\Lambda r^2}{d_1d_2}+\frac{m}{r^{d_3}}+\frac{Q^{d_1/2}d_3\ln{r}}{r^{d-3}}\big)\big]^{\frac{1}{4}}, &  & q=\frac{d_1}{4}.
  \end{array}\right.\label{39h}\end{split}
  \end{equation}
  For obtaining real solutions, we take $\Lambda>0$ and $\mu_4>0$. Since the spacetime dimension $d=9$ produces negative value for $\mu_4$ while other choices of $d$ lead to positive values, so it is convenient to ignore the case of $d=9$. Our numerical calculations show that for the determination of black hole solution, one needs to take $d>9$. In the limit $r\rightarrow\infty$, the metric function corresponding to pure quasi-topological-power-Yang-Mills solution tends to
  \begin{equation}
  f_p(r)=k\mp\bigg(\frac{2\Lambda}{\mu_4d_1d_2}\bigg)^{\frac{1}{4}}r^2.\label{40h}
  \end{equation}
  Note that, the minus and plus signs are defined respectively for $k=1$ and $k=-1$, whereas the other cases lead to naked singularity. Thus, the choice $\Lambda>0$ in this metric function may describe asymptotically AdS and dS pure quasi-topological black holes with $k=-1$ and $k=1$, respectively. These power-Yang-Mills black holes possess horizons, if the metric function satisfies the condition $f_p(r_+)=0$. Hence, on the basis of appropriate choices for the parameters $d$, $m$, $Q$, $\Lambda$ and $\mu_4$, the pure quasi-topological solution can describe a black hole having horizons. In this regard, we plot metric function (\ref{39h}) as a function of $r$ with $\Lambda=1$ in Figs. \ref{Asaif11}-\ref{Asaif14}. Those points for which the curve intersects the horizontal axis correspond to the location of horizons. It may be noted that for $k=-1$ and $d>9$, the solution (\ref{39h}) describes AdS black hole with two horizons, an extreme dS black hole and a naked singularity. It is also shown that the Yang-Mills charge $Q$ and nonlinearity parameter $q$ affect the horizon structure of the black hole. The case $q=1$ in Fig. \ref{Asaif13} corresponds to the behaviour of the black hole solution of quasi-topological gravity with Yang-Mills source. Fig. \ref{Asaif14} shows the behaviour of metric function (\ref{39h}) in those spacetime dimensions which satisfy the condition $q=d_1/4$. It can also be verified that for a positive value of quasi-topological parameter $\mu_4$, the Kretschmann scalar associated with pure quasi-topological solution in the vicinity of $r=0$ takes the form as
  \begin{equation}
  K\propto\big(\frac{m}{\mu_4}\big)^{1/2}r^{-(d_1)/2},\label{41h}
  \end{equation}
  which diverges at $r=0$. Therefore, the pure quasi-topological power-Yang-Mills black hole has an essential central singularity.
\section{Thermodynamics of pure quasi-topological power-Yang-Mills black holes}
In order to study thermodynamic properties of pure quasi-topological black holes described by metric function (\ref{39h}) we will work out various thermodynamic quantities. In this case too, the mass density follows from Eq. (\ref{24h}) and, as a function of outer horizon, it can be obtained as
 \begin{equation}\begin{split}
M&=\left\{ \begin{array}{rcl}
\frac{d_2}{16\pi}\bigg(\mu_4k^4r_+^{d_9}-\frac{2\Lambda r_+^{d_1}}{d_1d_2}-\frac{d_3^qd_2^{q-1}Q^{2q}r_+^{d_1-4q}}{d_1-4q}\bigg), & 
& q\neq\frac{d_1}{4}, \\\frac{d_2}{16\pi}\bigg(\mu_4k^4r_+^{d_9}-\frac{2\Lambda r_+^{d_1}}{d_1d_2}-d_3Q^{d_1/2}\ln{r_+}\bigg), &  & q=\frac{d_1}{4}.
\end{array}\right.\label{42h}\end{split}
\end{equation}
Using the condition $f_p(r_+)=0$ and the polynomial equation (\ref{38h}), it is straightforward to obtain Hawking temperature as
\begin{equation}\begin{split}
T_H&=\left\{ \begin{array}{rcl}
 \frac{1}{4\pi}\bigg[\frac{r_+^8}{4\mu_4k^3}\bigg(\frac{\mu_4k^4d_1}{r_+^9}-\frac{2\Lambda}{d_2r_+}-\frac{d_3^qd_2^{q-1}Q^{2q}}{r_+^{4q+1}}\bigg)-\frac{2k}{r_+}\bigg]& 
& q\neq\frac{d_1}{4}, \\\frac{1}{4\pi}\bigg[\frac{r_+^8}{4\mu_4k^3}\bigg(\frac{\mu_4k^4d_1}{r_+^9}-\frac{2\Lambda}{d_2r_+}-\frac{d_3Q^{d_1/2}}{r_+^{d}}\bigg)-\frac{2k}{r_+}\bigg], &  & q=\frac{d_1}{4}.
\end{array}\right.\label{43h}\end{split}
\end{equation}
\begin{figure}[h]
	\centering
	\includegraphics[width=0.8\textwidth]{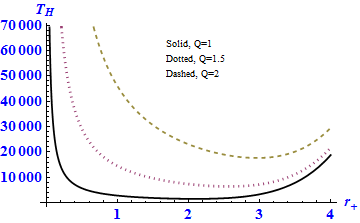}
	\caption{Dependence of temperature $T_H$ (Eq. (\ref{43h})) on the Yang-Mills charge $Q$ for fixed values of $d=11$, $q=2$, $k=-1$, $\mu_4=0.004$ and $\Lambda=1$.}\label{Asaif15}
\end{figure}
\begin{figure}[h]
	\centering
	\includegraphics[width=0.8\textwidth]{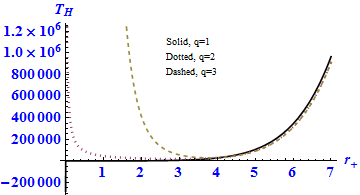}
	\caption{Dependence of temperature $T_H$ (Eq. (\ref{43h})) on the parameter $q$ for fixed values of $d=11$, $Q=2$, $k=-1$, $\mu_4=0.004$ and $\Lambda=1$.}\label{Asaif16}
\end{figure}
\begin{figure}[h]
	\centering
	\includegraphics[width=0.8\textwidth]{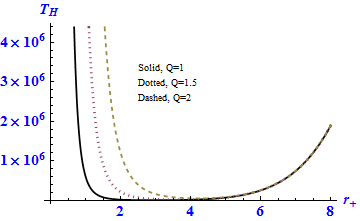}
	\caption{Dependence of temperature $T_H$ (Eq. (\ref{43h})) on the Yang-Mills charge $Q$ for fixed values of $d=13$, $q=d_1/4=3$, $k=-1$, $\mu_4=0.004$ and $\Lambda=1$.}\label{Asaif17}
\end{figure}

Fig. \ref{Asaif15} describes the behaviour of Hawking temperature as a function of the outer horizon for some values of $Q$. It can be clearly seen that for smaller black holes when Yang-Mills charge increases, the temperature also increases while for the larger ones the Hawking temperature does not change with this increase in $Q$. The choice $q=1$ in Fig. \ref{Asaif16} shows the temperature of Yang-Mills quasi-topological black holes. It can also be observed that for smaller black holes the nonlinearity parameter $q$ possesses greater influence on the temperature. However, for larger black holes, the effect of parameter $q$ is negligible. Similarly, Fig. \ref{Asaif17} shows the plot of Hawking temperature when the spacetime dimensions satisfy the case $q=d_1/4$. For this choice too, the Yang-Mills charge affects the temperature of quasi-topological black holes for small values of $r_+$ while for large horizon radii, the temperature although increases but the effect of $Q$ is negligible. 

Again, using the same technique as in Ref. \cite{59W}, the entropy density of pure quasi-topological black hole can be calculated as
\begin{equation}
\mathcal{S}=\frac{d_2k^3\mu_4}{d_8}r_+^8.\label{44h} 
\end{equation}
Our calculations show that the temperature (\ref{43h}) is equal to $\bigg(\frac{\partial M}{\partial \mathcal{S}}\bigg)$. Thus, the power-Yang-Mills black holes in pure quasi-topological gravity also satisfy the first law of thermodynamics (\ref{30h}) if the power-Yang-Mills potential associated with (\ref{39h}) is given by
 \begin{equation}\begin{split}
 \mathcal{U}=\frac{\partial M}{\partial\mathcal{S}}&=\left\{ \begin{array}{rcl}
 -\frac{q d_3^qd_2^qQ^{2q-1}r_+^{d_1-4q}}{8\pi(d_1-4q)}, & 
 & q\neq\frac{d_1}{4}, \\-\frac{d_1d_2d_3}{32\pi}Q^{d_3/2}\ln{r_+}, &  & q=\frac{d_1}{4}.
 \end{array}\right.\label{45h}\end{split}
 \end{equation}
 The heat capacity can be obtained as
  \begin{equation}\begin{split}
  C_H=T_H\bigg(\frac{\partial \mathcal{S}}{\partial T_H}\bigg)_{\tilde{Q}}&=\left\{ \begin{array}{rcl}
  \frac{d_2k^3\mu_4 r_+^{d_8}\big(d_2k^4\mu_4d_9r_+^{4q-8}-2\Lambda r_+^{4q}-d_3^qd_2^qQ^{2q}\big)}{\big((4q-7)d_2^qd_3^qQ^{2q}-14\Lambda r_+^{4q}-\mu_4k^4d_2d_9r_+^{4q-8}\big)}, & 
  & q\neq\frac{d_1}{4}, \\\frac{d_2k^3\mu_4 r_+^{d_8}\big(\mu_4k^4d_2d_9r_+^{d_9}-2\Lambda r_+^{d_1}-d_2d_3Q^{d_1/2}\big)}{\big(Q^{d_1/2}d_2d_3d_8-14\Lambda r_+^{d_1}-k^4\mu_4d_2d_9r_+^{d_9}\big)}, &  & q=\frac{d_1}{4}.
  \end{array}\right.\label{46h}\end{split}
  \end{equation}
  \begin{figure}[h]
  	\centering
  	\includegraphics[width=0.8\textwidth]{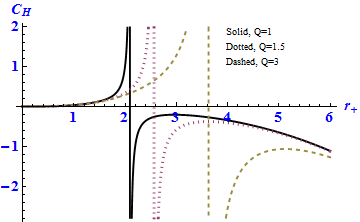}
  	\caption{Dependence of heat capacity $C_H$ (Eq. (\ref{46h})) on the parameter $\tilde{Q}=Q/4\pi$ for fixed values of $d=11$, $q=2$, $k=-1$, $\mu_4=0.004$ and $\Lambda=1$.}\label{Asaif18}
  \end{figure}
\begin{figure}[h]
	\centering
	\includegraphics[width=0.8\textwidth]{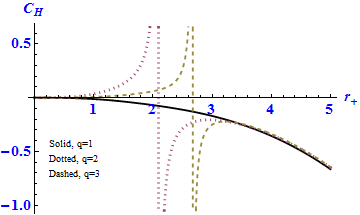}
	\caption{Plot of heat capacity $C_H$ (Eq. (\ref{46h})) for different values of the nonlinearity parameter $q$ and fixed values of $d=11$, $Q=2$, $k=-1$, $\mu_4=0.004$ and $\Lambda=1$.}\label{Asaif19}
\end{figure}
\begin{figure}[h]
	\centering
	\includegraphics[width=0.8\textwidth]{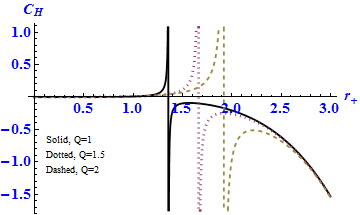}
	\caption{Dependence of heat capacity $C_H$ (Eq. (\ref{46h})) on the parameter $\tilde{Q}=Q/4\pi$ for fixed values of $d=13$, $q=d_1/4=3$, $k=-1$, $\mu_4=0.004$ and $\Lambda=1$.}\label{Asaif20}
\end{figure}
  Note that, the parameter $Q$ is related to Yang-Mills charge $\tilde{Q}$ through Eq. (\ref{26h}). The above expression reduces to the heat capacity of Yang-Mills black holes in pure quasi-topological gravity \cite{31W} when $q=1$. One can also observe that for positive values of $\mu_4$ and $\Lambda$, the above heat capacity is negative when $q=1$. Hence, it follows that the Yang-Mills black holes are thermodynamically unstable in the canonical ensemble. However, the local stability of power-Yang-Mills black holes (i.e. when $q\neq 1$) can be described from the plot of heat capacity as a function of $r_+$. Fig. \ref{Asaif18} shows the plot of heat capacity for different values of $Q$ and fixed values of other parameters involved in (\ref{46h}). The region in which this thermodynamic quantity is positive corresponds to local stability while its negativity implies local instability of pure quasi-topological black holes. Fig. \ref{Asaif19} depicts the behaviour of this quantity for different values of the parameter $q$. One can clearly see that the Yang-Mills black hole (i.e. when $q=1$) is unstable in the canonical ensemble. However, for $q\neq 1$, there exist regions of local stability in this ensemble. Hence, the nonlinearity of Yang-Mills field induces certain affects on the local thermodynamic stability of black holes. Finally, the local stability of pure quasi-topological black holes in spacetime dimensions satisfying $q=d_1/4$ can be examined from Fig. \ref{Asaif20}. One can also analyze the possibility of first and second order transitions from these plots of heat capacity. The points at which heat capacity vanishes correspond to first order transitions of black hole. The points which make this quantity divergent imply the possibility of second order phase transitions.
  
  In the grand canonical ensemble, the entropy $\mathcal{S}$ and charge $\tilde{Q}$ should be considered as variables. In this ensemble, the local thermodynamic stability can be determined from the positivity of both $\big(\partial^2M/\partial\tilde{Q}^2\big)$ and the determinant of the Hessian matrix. Using the above value of mass in (\ref{42h}) we can compute
  \begin{equation}\begin{split}
  \bigg(\frac{\partial^2 M}{\partial Q^2}\bigg)&=\left\{ \begin{array}{rcl}
 -\frac{q(2q-1)d_3^qd_2^q(4\pi)^{2q}\tilde{Q}^{2q-2}}{8\pi (d_1-4q)}r_+^{d_1-4q}, & 
  & q\neq\frac{d_1}{4}, \\-\frac{d_1d_2d_3^2 (4\pi)^{2q}\tilde{Q}^{d_5/2}}{64}\ln{r_+}, &  & q=\frac{d_1}{4}.
  \end{array}\right.\label{47h}\end{split}
  \end{equation}
The above quantity is negative when $q=1$ and $q\leq d_1/4$. Thus, for these two choices the black hole is unstable in this ensemble. However, for $q>d_1/4$, the above quantity is positive and so we need to check the behaviour of the Hessian matrix determinant for the investigation of thermal stability. The determinant of Hessian matrix in terms of the outer horizon can be computed as

  \begin{equation}\begin{split}
  det \textbf{H}&=\left\{ \begin{array}{rcl}
  \frac{q(2q-1)d_3^qd_2^q(4\pi)^{2q}\tilde{Q}^{2q-2}r_+^{d_1-4q}}{32\pi^2 d_2\mu_4k^3(d_1-4q)}A_p(r_+)-\frac{4q^2d_3^{2q}d_2^{2q-2}(4\pi)^{2q}\tilde{Q}^{4q-2}}{256\pi^2\mu_4^2k^6r_+^{8q-14}}, & 
  & q>\frac{d_1}{4}, \\ \frac{d_1d_2d_3^2(4\pi)^{d_1/2}\tilde{Q}^{d_1/2}\ln{r_+}}{256\pi^2d_2\mu_4k^3}A_p(r_+)-\frac{d_1^2(4\pi)^{d_1/2}\tilde{Q}^{d_3}d_3^2}{1624\pi^2\mu_4^2k^6r_+^{2d_8}}, &  & q=\frac{d_1}{4},  \end{array}\right.\label{48h}\end{split}
  \end{equation}
  where 
   \begin{equation}\begin{split}
  A_p(r_+)&=\left\{ \begin{array}{rcl}
  \frac{kd_9}{4r_+^{d_7}}+\frac{7\Lambda}{2\mu_4d_2k^3r_+^{d_{15}}}+\frac{(7q-4)d_3^qd_2^{q-1}(4\pi)^{2q}\tilde{Q}^{2q}}{4\mu_4k^3r_+^{d_{15}+4q}}, & 
  & q>\frac{d_1}{4}, \\ \frac{kd_9}{r_+^{d_7}}+\frac{7\Lambda}{2\mu_4d_2k^3r_+^{d_{15}}}-\frac{d_3d_8(4\pi)^{d_1/2}\tilde{Q}^{d_1/2}}{4\mu_4k^3r_+^{2d_8}}, &  & q=\frac{d_1}{4}.
  \end{array}\right.\label{49h}\end{split}
  \end{equation}
    \begin{figure}[h]
  	\centering
  	\includegraphics[width=0.8\textwidth]{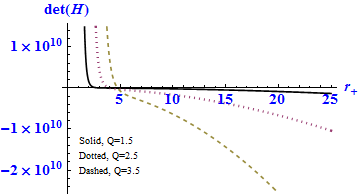}
  	\caption{Dependence of the determinant $det\textbf{H}$ (Eq. (\ref{48h})) on the parameter $Q$ for fixed values of $d=11$, $q=3$, $k=-1$, $\mu_2=-0.06$, $\mu_3=-0.1$, $\mu_4=0.03$ and $\Lambda=1$.}\label{AAli4}
  \end{figure}
\begin{figure}[h]
	\centering
	\includegraphics[width=0.8\textwidth]{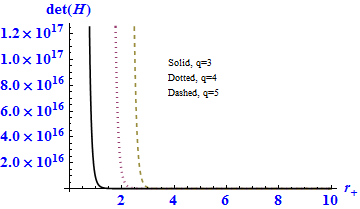}
	\caption{Plot of the determinant $det \textbf{H}$ (Eq. (\ref{48h})) for different values of parameter $q$ and fixed values of $d=11$, $Q=3$, $k=-1$, $\mu_2=-0.06$, $\mu_3=-0.1$, $\mu_4=0.03$ and $\Lambda=1$.}\label{AAli5}
\end{figure}
The plot of $det\textbf{H}$ for various values of charge $\tilde{Q}=Q/4\pi$ is shown in Fig. \ref{AAli4}. One can see that there exists a value $r_{0}$ such that $det\textbf{H}$ is positive when $r_+<r_{0}$. This indicates the region of black hole's stability. However, as $r_+$ increases its value from $r_0$, this determinant becomes negative and we have thermodynamic instability. Similarly, the behaviour of this determinant for different values of nonlinearity parameter $q$ is given in Fig. \ref{AAli5}. One can observe that as the parameter $q$ increases, the horizon radius of stable black hole also increases. Hence, we conclude that unlike Yang-Mills black holes \cite{31W} of pure quasi-topological gravity, the power-Yang-Mills black holes in this theory could be thermodynamically stable in the grand canonical ensemble.

\section{Thermodynamics of quartic quasi-topological rotating black branes with power-Yang-Mills source}
\begin{figure}[h]
 	\centering
 	\includegraphics[width=0.8\textwidth]{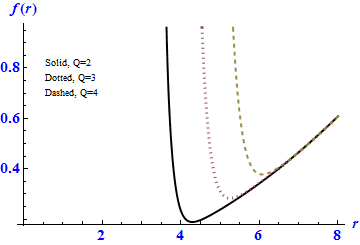}
 	\caption{Dependence of the solution $f(r)$ (Eq. (\ref{16h})) on the parameter $Q$ for fixed values of $d=11$, $m=1$, $q=7$, $k=0$, $\mu_2=-0.06$, $\mu_3=-0.1$, $\mu_4=2^{10}$ and $\Lambda=-1$.}\label{Asaif21}
 \end{figure}

\begin{figure}[h]
	\centering
	\includegraphics[width=0.8\textwidth]{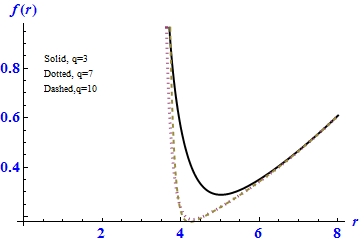}
	\caption{Plot of the solution $f(r)$ (Eq. (\ref{16h})) for different values of the parameter $q$ and fixed values of $d=11$, $m=1$, $Q=3$, $k=0$, $\mu_2=-0.06$, $\mu_3=-0.1$, $\mu_4=2^{10}$ and $\Lambda=-1$.}\label{Asaif22}
\end{figure}

\begin{figure}[h]
  	\centering
  	\includegraphics[width=0.8\textwidth]{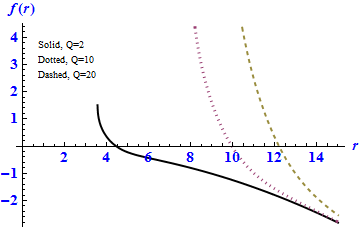}
  	\caption{Dependence of the solution $f(r)$ (Eq. (\ref{16h})) on the parameter $Q$ for fixed values of $d=11$, $m=1$, $q=7$, $k=0$, $\mu_2=-0.06$, $\mu_3=-0.1$, $\mu_4=2^{10}$ and $\Lambda=1$.}\label{Asaif23}
  \end{figure}

\begin{figure}[h]
	\centering
	\includegraphics[width=0.8\textwidth]{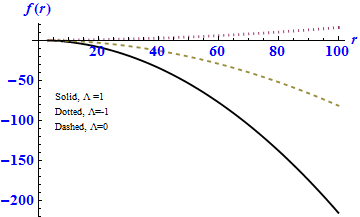}
	\caption{Plot of the solution $f(r)$ (Eq. (\ref{16h})) for different values of the cosmological constant and fixed values of $d=13$, $q=d_1/4=3$, $Q=4$, $k=0$, $\mu_2=-0.06$, $\mu_3=-0.1$ and $\mu_4=2^{10}$}. \label{Asaif24}
\end{figure}

In this section we will implement solution (\ref{16h}) for $k=0$ with a global rotation. This can be done, if we use the transformation describing the rotation boost in the $t-\phi_i$ planes, i.e.,
\begin{align}\begin{split}
&t\longmapsto\Xi t-\sum_{i=1}^{p}a_i\phi_i, \\& \phi_i\longmapsto\Xi\phi_i-\frac{a_i}{l^2}t.\label{50h}\end{split}
\end{align}
 The $SO(d_1)$ rotation group in $d$-dimensions contains the maximum number of rotational parameters. Hence, the independent parameters of rotation are of number $[d_1/2]$, where $[...]$ stands for the integer part. Thus, the line element for the rotating spacetime with flat horizon and $p\leq[d_1/2]$ rotation parameters can be given as
 \begin{equation}\begin{split}
 ds^2=&-f(r)\big(\Xi dt-\sum_{i=1}^{p}a_id\phi_i\big)^2+\frac{dr^2}{f(r)}+\frac{r^2}{l^4}\sum_{i=1}^{p}\big(a_idt-\Xi l^2d\phi_i\big)^2\\&-\frac{r^2}{l^2}\sum_{i<j}^{p}\big(a_id\phi_j-a_jd\phi_i\big)^2+r^2\sum_{i=1}^{d_2-p}dx_i^2,\label{51h}\end{split}
 \end{equation}
 where $\Xi=\sqrt{1+\sum_{i=1}^{p}a_i^2/l^2}$, $l$ is a scale factor related to cosmological constant and $a_i$'s are the $p$ parameters of rotation. It should be noted that static line element (\ref{8h}) and rotating metric (\ref{51h}) can be mapped locally onto each other, not globally. In order to study the physical properties of the solutions obtained for $k=0$ in quartic quasi-topological gravity coupled to power-Yang-Mills theory, we plot the metric function $f(r)$ for suitable values of parameters involved in it. In Figs. \ref{Asaif21}-\ref{Asaif24}, it is clear that the metric function possesses divergences at the central position $r=0$. One can also verify that the Kretschmann scalar diverges at this point. However, as $r$ becomes larger and larger, the behaviour of $f(r)$ is dependent on the value of the cosmological constant $\Lambda$. Hence, we observe from these graphs that the metric function approaches towards $+\infty$ when $\Lambda<0$ while it tends to $-\infty$ when $\Lambda>0$. Moreover, Fig. \ref{Asaif21} shows the behaviour of the metric function for different values of parameter $Q$ in AdS spacetime. It may be noted from here, that for fixed values of other parameters there exists a value $Q_{ext}$ for which we have an extreme black brane, while for the case $Q<Q_{ext}$, there may be black brane having two horizons and for $Q>Q_{ext}$ we will arrive at a naked singularity. The dependence of nonlinearity parameter $q$ on the metric function with negative cosmological constant can be visualized from Fig. \ref{Asaif22}. One can see that this parameter is also affecting the values of the horizons. The solution in dS spacetimes has also been plotted in Fig. \ref{Asaif23} for different values of Yang-Mils charge $Q$. Similarly, the behaviour of $f(r)$ for all the three cases of the cosmological constant has been compared in Fig. \ref{Asaif24}, when the spacetime dimensions satisfy $q=d_1/4$.        
Now, the Killing vector associated with the rotating black brane metric can be defined as
\begin{equation}
\mathcal{X}=\partial_t+\sum_{j=1}^{p}\Omega_j\partial_{\phi_j},\label{52h}
\end{equation}
where $\Omega_j$ refers to the angular velocity and is given by
\begin{equation}
\Omega_j=-\bigg(\frac{g_{t\phi_j}}{g_{\phi_j\phi_j}}\bigg)=\frac{a_j}{\Xi l^2}.\label{53h}
\end{equation}
The expression for Hawking temperature associated with this rotating black brane takes the form
\begin{equation}
T_H(r_+)=\frac{f'(r_+)}{4\pi \Xi}=\frac{r_+^2}{4\pi \Xi }\Upsilon'(r_+).\label{54h}
\end{equation}
 During the computation of thermodynamic quantities from the variation of action (\ref{2h}) with respect to metric tensor, one gets a total derivative surface term containing the derivatives of $\delta g_{\mu\nu}$ normal to the boundary. Since these derivative terms do not cancel with each other, so, the variation of the action is not well-defined. To handle this issue, it is convenient to add the Gibbons-Hawking surface term $\mathcal{I}_{b}$ with the bulk action (\ref{2h}). Thus, the variational principle would be well-defined if this boundary term can be included in the following form
 \begin{equation}
 \mathcal{I}_b=\mathcal{I}_b^{(I)}+\mathcal{I}^{(II)}+\mathcal{I}^{(III)}+\mathcal{I}_b^{(IV)},\label{55h}
 \end{equation} 
 where $\mathcal{I}_b^{(I)}$, $\mathcal{I}_b^{(II)}$, $\mathcal{I}_b^{(III)}$ and $\mathcal{I}_b^{(IV)}$ respectively, stand for the surface terms corresponding to Einstein \cite{62W}, second order Lovelock (Gauss-Bonnet) \cite{5W,63W}, cubic quasi-topological \cite{64W} and quartic quasi-topological \cite{65W} gravities. These terms are obtained as
 \begin{equation}
 \mathcal{I}_b^{(I)}=\frac{1}{8\pi}\int_{\partial \mathcal{M}}d^{d_1}x\sqrt{-\gamma}\mathcal{K}, \label{56h}
 \end{equation} 
 \begin{equation}
 \mathcal{I}_b^{(II)}=\frac{1}{8\pi}\int_{\partial \mathcal{M}}d^{d_1}x\sqrt{-\gamma}\frac{2\tilde{\mu}_2l^2}{3d_3d_4}\big(3\mathcal{K}\mathcal{K}_{ad}\mathcal{K}^{ad}-2\mathcal{K}_{ac}\mathcal{K}^{cd}\mathcal{K}^{a}_{d}-\mathcal{K}^3\big), \label{57h}
 \end{equation}  
 \begin{equation}\begin{split}
 \mathcal{I}_b^{(III)}=&\frac{1}{8\pi}\int_{\partial \mathcal{M}}d^{d_1}x\sqrt{-\gamma}\bigg[\frac{3\tilde{\mu}_3l^4}{5d_1d_2^2d_3d_6}\bigg(d_1\mathcal{K}^5-2\mathcal{K}^3\mathcal{K}_{ad}\mathcal{K}^{ad}+4d_2\mathcal{K}_{ab}\mathcal{K}^{ab}\mathcal{K}_{cd}\mathcal{K}^{d}_{e}\mathcal{K}^{ec}\\&-(5d_1-6)\mathcal{K}\mathcal{K}_{ab}\big(d_1\mathcal{K}^{ab}\mathcal{K}^{cd}\mathcal{K}_{cd}-d_2\mathcal{K}^{ac}\mathcal{K}^{bd}\mathcal{K}_{cd}\big)\bigg)\bigg], \label{58h}\end{split}
 \end{equation}  
 and
  \begin{equation}\begin{split}
  \mathcal{I}_b^{(IV)}=&\frac{1}{8\pi}\int_{\partial \mathcal{M}}d^{d_1}x\sqrt{-\gamma}\bigg[\frac{2\tilde{\mu}_4l^6}{7d_1d_2d_3d_8(d_1^2-3d_1+3)}\bigg(\alpha_1\mathcal{K}^3\mathcal{K}^{ab}\mathcal{K}_{ac}\mathcal{K}_{bd}\mathcal{K}^{cd}\\&+\alpha_2\mathcal{K}^2\mathcal{K}^{ab}\mathcal{K}_{ab}\mathcal{K}^{cd}\mathcal{K}^{e}_{c}\mathcal{K}_{de}+\alpha_3\mathcal{K}^2\mathcal{K}^{ab}\mathcal{K}_{ac}\mathcal{K}_{bd}\mathcal{K}^{ce}\mathcal{K}^{d}_{e}\\&+\alpha_4\mathcal{K}\mathcal{K}^{ab}\mathcal{K}_{ab}\mathcal{K}^{cd}\mathcal{K}^{e}_{c}\mathcal{K}^{f}_{d}\mathcal{K}_{ef}+\alpha_5\mathcal{K}\mathcal{K}^{ab}\mathcal{K}^{c}_{a}\mathcal{K}_{bc}\mathcal{K}^{de}\mathcal{K}^{f}_{d}\mathcal{K}_{ef}\\&+\alpha_6\mathcal{K}\mathcal{K}^{ab}\mathcal{K}_{ac}\mathcal{K}_{bd}\mathcal{K}^{ce}\mathcal{K}^{df}\mathcal{K}_{ef}+\alpha_7\mathcal{K}^{ab}\mathcal{K}^{c}_{a}\mathcal{K}_{bc}\mathcal{K}^{de}\mathcal{K}_{df}\mathcal{K}_{eg}\mathcal{K}^{fg}\bigg)\bigg], \label{59h}\end{split}
  \end{equation}
  where $\gamma_{ab}$ refers to the induced metric tensor on the boundary $\partial\mathcal{M}$ while $\mathcal{K}$ stands for the trace of the extrinsic curvature $\mathcal{K}^{ab}$ of this boundary. It is worthwhile to note that the value of the total action $\mathcal{I}_{bulk}+\mathcal{I}_b$ is infinite on the solutions. However, this divergence can be removed with the help of the counter-term method \cite{66W,67W,68W,69W,70W}. The counter-term action needed for the removal of this divergence can be written as
  \begin{equation}
  \mathcal{I}_{count}=-\frac{1}{8\pi}\int_{\partial\mathcal{M}}d^{d_1}x\sqrt{-\gamma}\bigg(\frac{d_2}{L_{eff}}\bigg),\label{60h}
  \end{equation} 
   where $L_{eff}$ describes the effective scale length factor related to $l$ and parameters $\tilde{\mu}_2$, $\tilde{\mu}_3$ and $\tilde{\mu}_4$. Note that, $L_{eff}$ reduces to $l$ when the coupling constants i.e. $\tilde{\mu}_2$, $\tilde{\mu}_3$ and $\tilde{\mu}_4$ approach zero. Using the counter-term method, the overall action $\mathcal{I}_{bulk}+\mathcal{I}_b+\mathcal{I}_{count}$ becomes finite and can be used to compute the conserved and thermodynamic quantities. 
   
   The conserved quantities related to the timelike $\partial_t$ and rotational $\partial_{\phi_j}$ Killing vector fields can be computed as
   \begin{equation}
   M=\frac{(d_1\Xi^2-1)m}{16\pi d_2l^{d_2}}, \label{61h}
   \end{equation} 
    \begin{equation}
   J_i=\frac{d_1\Xi a_im}{16\pi d_2l^{d_2}}. \label{62h}
   \end{equation} 
   One can clearly see that by choosing the rotation parameter $a_i$ equal to zero or $\Xi=1$, the angular momentum $\textbf{J}$ vanishes and (\ref{61h}) then describes the mass of the static black hole. The Yang-Mills charge per unit volume $\mathcal{V}_{d_2}$ in this case can be obtained as
   \begin{eqnarray}
   \tilde{Q}=\frac{\Xi Q}{4\pi l^{d_4}}.\label{63h}
   \end{eqnarray}
   It is well-known that entropy is the quarter of horizon area \cite{71W,72W,73W}. Using this, the entropy density for the power-Yang-Mills black brane can be obtained as
   \begin{equation}
   \mathcal{S}=\frac{\Xi r_+^{d_2}}{4l^{d_4}}.\label{64h}
   \end{equation} 
  In order to check the validity of the first law, it is more convenient to calculate the mass in terms of extensive variables $\mathcal{S}$, $\tilde{Q}$ and $\textbf{J}$. Therefore, by taking $\mathcal{Z}=\Xi^2$ and using Eqs. (\ref{61h})-(\ref{62h}), we construct the Smarr-type formula as
  \begin{equation}
  M(\mathcal{S},\tilde{Q},J)=\frac{(d_1\mathcal{Z}-1)J}{d_1l\sqrt{\mathcal{Z}(\mathcal{Z}-1)}}.\label{65h}
  \end{equation}
  It should be noted that the parameter $\mathcal{Z}$ should be dependent on the extensive parameters. Using Eqs. (\ref{63h})-(\ref{64h}) and the condition for event horizon i.e. $f(r_+)=0$, it is possible to obtain an equation $\mathcal{E}(\mathcal{S},\tilde{Q},J)=0$, whose positive real root is $\mathcal{Z}=\Xi^2$ and
     \begin{equation}\begin{split}
   \mathcal{E}(\mathcal{S},\tilde{Q},J) &=\left\{ \begin{array}{rcl}
   \frac{16\pi l^{d_3}d_2J\mathcal{Z}^{d_1/2d_2}}{d_1\sqrt{\mathcal{Z}(\mathcal{Z}-1)}}+\frac{2\Lambda l^{d_1d_4/d_2}(4)^{d_1/d_2}}{d_1d_2\mathcal{S}^{-d_1/d_2}}+\frac{(\pi\tilde{Q})^{2q}d_2^{q-1}d_3^q\big(4l^{d_4}\big)^{(d_1+2qd_4)/d_2}}{(d_1-4q)\mathcal{Z}^{d_4q/d_2}\mathcal{S}^{(4q-d_1)/d_2}}, & 
    & q\neq\frac{d_1}{4}, \\ \frac{16\pi l^{d_3}d_2J\mathcal{Z}^{d_1/2d_2}}{d_1\sqrt{\mathcal{Z}(\mathcal{Z}-1)}}+\frac{2\Lambda l^{d_1d_4/d_2}(4)^{d_1/d_2}}{d_1d_2\mathcal{S}^{-d_1/d_2}}+\frac{\big(4l^{d_4}\pi\tilde{Q}\big)^{d_1/d_2}d_3}{d_2\mathcal{Z}^{d_1/4}}\ln{\bigg(\frac{4l^{d_4}S}{\mathcal{Z}^{1/2}}\bigg)}, &  & q=\frac{d_1}{4}.  \end{array}\right.\label{66h}\end{split}
    \end{equation}
    Now, it is straightforward to write the mass $M(\mathcal{S},\tilde{Q},J)$ in terms of the extensive parameters and compute the intensive parameters conjugate to them as follows
    \begin{align}\begin{split}
    &T_H=\bigg(\frac{\partial M}{\partial \mathcal{S}}\bigg)_{J,\tilde{Q}},\\&\Omega_k=\bigg(\frac{\partial M}{\partial J_k}\bigg)_{\mathcal{S},\tilde{Q}},\label{67h}\end{split}
    \end{align}
    while, the power-Yang-Mills potential is given by
     \begin{equation}\begin{split}
    \mathcal{U}=&\bigg(\frac{\partial M}{\partial \tilde{Q}}\bigg)_{\mathcal{S}, J}=\left\{ \begin{array}{rcl}
    \frac{2qJ(d_3\mathcal{Z}+1)d_2^{q-1}d_3^q(\pi\tilde{Q})^{2q}\big(4l^{d_4}\big)^{(2qd_4+1)/d_2}\mathcal{S}^{(d_1-4q)/d_2}}{2d_1l(d_1-4q)\mathcal{Z}^{qd_4/d_2}(\mathcal{Z}(\mathcal{Z}-1))^{3/2}\mathfrak{Y}(\mathcal{S},\tilde{Q},J)}, & 
    & q\neq\frac{d_1}{4}, \\\frac{d_3(d_3\mathcal{Z}+1)(4\pi)^{d_1/d_2}l^{d_1d_4/d_2}J\tilde{Q}^{1/d_2}}{2ld_2^2\mathcal{Z}^{d_1/4}(\mathcal{Z}(\mathcal{Z}-1))^{3/2}\mathfrak{Y}(\mathcal{S},\tilde{Q},J)}\ln{\big(\frac{4\mathcal{S}l^{d_4}}{\sqrt{\mathcal{Z}}}\big)}, &  & q=\frac{d_1}{4},  \end{array}\right.\label{68h}\end{split}
    \end{equation}
    in which
    \begin{equation}\begin{split}
    \mathfrak{Y}(\mathcal{S},\tilde{Q},J)=&\left\{ \begin{array}{rcl}
    \frac{q(\pi\tilde{Q})^{2q}d_2^qd_3^qd_4\big(4l^{d_4}\big)^{(d_1+2qd_4)/d_2}\mathcal{S}^{(d_1-4q)/d_2}}{(d_1-4q)\mathcal{Z}^{1+d_4q/d_2}}+\frac{8\pi l^{d_3}J(d_3\mathcal{Z}+1)\mathcal{Z}^{d_1/2d_2}}{d_1(\mathcal{Z}(\mathcal{Z}-1))^{3/2}}, & 
    & q\neq\frac{d_1}{4}, \\\frac{8\pi l^{d_3}J(d_3\mathcal{Z}+1)\mathcal{Z}^{d_1/2d_2}}{d_1(\mathcal{Z}(\mathcal{Z}-1))^{3/2}}+\frac{d_3(4\pi\tilde{Q})^{d_1/d_2}l^{d_1d_4/d_2}}{2d_2\mathcal{Z}^{1+(d_1/4)}}\bigg(1+\frac{d_1}{2}\ln{\big(\frac{4\mathcal{S}l^{d_4}}{\sqrt{\mathcal{Z}}}\big)}\bigg), &  & q=\frac{d_1}{4}.  \end{array}\right.\label{69h}\end{split}
    \end{equation}
   Our calculations showed that the angular velocity and Hawking temperature in (\ref{67h}) are same as (\ref{53h}) and (\ref{54h}), respectively. Thus, our power-Yang-Mills rotating black brane satisfy the first law as
   \begin{equation}
   dM=T_HdS+\sum_{i=1}^{p}\Omega_idJ_i+\mathcal{U}d\tilde{Q}.\label{70h}
   \end{equation} 
  
\section{Concluding Remarks} 

In this work, we mainly focused on the physical and thermodynamic properties of quartic quasi-topological black holes with power-Yang-Mills source. First, we have considered the fourth order quasi-topological gravity and coupled it with the power-Yang-Mills theory. From the Wu-Yang ansatz, the gauge potentials are defined and the gravitational field equations are solved. In this context, two analytic power-Yang-Mills black hole solutions are derived for $\mu_4>0$ and $\mu_4<0$ in this theory. It is shown that the real solutions exist only when $\mu_4>0$. We have also write the two expressions separately for the metric function valid in spacetime dimensions when $q\neq d_1/4$ and $q=d_1/4$. We also studied the physical properties of these black holes and plot the associated solution $f(r)$ given in (\ref{16h}) for various values of the parameters $m$, $q$, $Q$, $\Lambda$ and $\mu_i$'s. Depending on the suitable choices for these parameters, either the solution describes a black hole which can possesses one or more horizons or it can describes a naked singularity. It is shown that variations in quasi-topological parameter $\mu_4$ affect the position of the horizon. Similarly, it can also be concluded that the value of the outer horizon is not affected by the charge parameter $Q$, however, the inner horizon increases with the increases of Yang-Mills magnetic charge. It should be noted that for $q=1$, the solution (\ref{16h}) yields the metric function of Yang-Mills black hole \cite{31W} in this theory. In addition to this, we have also studied thermodynamics of these power-Yang-Mills black holes. During this study, we worked out different thermodynamic quantities and plotted them as well. The region where positive temperature arises implies that the black hole is physical. It is also shown that the AdS power-Yang-Mills black holes may have larger range of parameters with positive temperature than the dS black holes. In the canonical ensemble, the positivity of specific heat capacity implies local thermodynamic stability. Hence, from the plots of heat capacity we have concluded that the outer horizon of stable power-Yang-Mills black holes increases with the increase in charge $Q$. The effects of nonlinearity parameter $q$ on the stability of black holes in this ensemble have also been shown. One can also observe that the stable power-Yang-Mills black holes have larger outer horizons than the Yang-Mills black holes. Thermodynamic stability in grand canonical ensemble have also been investigated. It is shown that the black holes with $q\leq d_1/4$ are unstable in this ensemble. However, for $q>d_1/4$ there exist stability regions for black holes. From the plots of the Hessian matrix, we have concluded that the smaller black holes could be stable in this ensemble. However, when the outer horizon $r_+$ increases then $det\textbf{H}$ is negative and so we have thermodynamic instability of the black holes. Thus, it can be concluded that the power-Yang-Mills field produces the possibility for local stability in the grand canonical ensemble. This behaviour is in contrast to the Yang-Mills theory \cite{31W}, where quasi-topological black holes are locally unstable in this ensemble.

 In addition to quartic quasi-topological black holes, we also derived a new family of black hole solutions in pure quasi-topological theory within the framework of power-Yang-Mills sources. The associated plots of pure quasi-topological black hole solution (\ref{39h}) for suitable values of parameters show that for $k=-1$ and $d>9$, this solution describes AdS black hole with two horizons, an extreme dS black hole and a naked singularity. It is shown that by choosing $\Lambda>0$, the asymptotic expression of metric function (\ref{40h}) describes the asymptotically AdS and dS power-Yang-Mills black holes for $k=-1$ and $k=1$, respectively. The effects of Yang-Mills charge and parameter $q$ on the horizon structure of black holes can also be observed from these plots. Note that, the case $q=1$ in (\ref{39h}) gives the black hole solution of pure quasi-topological gravity with Yang-Mills source. The local thermodynamic stability in both canonical and grand canonical ensembles have also been probed. Our results show that there exist the regions of local stability for the power-Yang-Mills black holes in the canonical ensemble. This can be seen from the corresponding plots of heat capacity and Hawking temperature. Moreover, from the plots of the determinant of Hessian matrix, one can identify the regions of local stability for these black holes in the grand canonical ensemble. It should be noted that it is the nonlinearity of Yang-Mills field that plays the main role in the stability of black holes. For $q=1$, our results correspond to those of the Yang-Mills black holes in pure quasi-topological gravity which are unstable in both canonical and grand canonical ensembles.  
 
 In the last part of our paper, we have assumed a general rotating line element with $p\leq[(d-1)/2]$ rotational parameters and study the rotating black branes of quartic quasi-topological gravity coupled to power-Yang-Mills theory. The plots of metric function (\ref{16h}) with $k=0$ show that for fixed values of parameters $d$, $q$, $m$, $\mu_2$, $\mu_3$ and $\mu_{4}$, it can describe the power-Yang-Mills black brane with inner and outer horizons for $Q<Q_{ext}$, extremal black brane for $Q=Q_{ext}$ and naked singularity for $Q>Q_{ext}$. These plots of the metric function show the effects of Yang-Mills charge $Q$ and nonlinearity parameter $q$ in all the three cases for cosmological constant $\Lambda$. Here, we included the generalized Gibbons-Hawking surface terms for the quasi-topological gravity which made the action well-defined. We calculated the Hawking temperature and angular velocities by taking the analytic continuation of the metric. In order to derive the finite action and conserved quantities, we have used the counter-term method. It is shown that the conserved quantities of these power-Yang-Mills black branes are independent of the coupling coefficients i.e. $\mu_2$, $\mu_3$ and $\mu_{4}$ for fixed values of mass, Yang-Mills charge and rotation parameters $a_i$. However, one can notice that the thermodynamic quantities depend indirectly on these coupling coefficients through the value of the outer horizon $r_+$. Furthermore, we have also derived the Smarr-type formula which describes the mass density in terms of entropy, Yang-Mills magnetic charge and angular momenta. We showed that the first law is satisfied for the power-Yang-Mills black branes obtained in this paper. It is worthwhile to note that for $q=1$, these results correspond to the Yang-Mills black branes of quasi-topological gravity.
  
 It would be very interesting to study the power-Yang-Mills black holes in quintic quasi-topological gravity. In addition to this, the study of black holes and black branes of quasi-topological gravity coupled to Yang-Mills theory in Lifshitz spacetime \cite{20W} could also be  very interesting.

 \appendix 
 
 \section{Appendix} 
 
The coupling coefficients $b_i$'s and $c_i$'s introduced in the Lagrangians (Eqs. (\ref{5h})-(\ref{6h})) of the cubic and quartic quasi-topological gravities are, respectively, defined as follows:
\begin{align}\begin{split}
&b_1=9d_1-15,\\& b_2=-24d_2,\\& b_3=24d,\\&b_4=48d_2,\\&b_5=-12(3d_1-1),\\&b_6=3d.\label{71h}\end{split}
\end{align}
\begin{align}\begin{split}
&c_1=-d_2(d_1^7-3d_1^6-29d_1^5+170d_1^4-349d_1^3+348d_1^2-180d_1+36),\\& c_2=-4d_4(2d_1^6-20d_1^5+65d_1^4-81d_1^3+13d_1^2+45d_1-18),\\& c_3=-64d_2(3d_1^2-8d_1+3)(d_1^2-3d_1+3),\\&c_4=-(d_1^8-6d_1^7+12d_1^6-22d_1^5+114d_1^4-345d_1^3+468d_1^2-270d_1+54),\\&c_5=16d_2(10d_1^4-51d_1^3+93d_1^2-72d_1+18),\\&c_6=-32d_2^2d_4^2(3d_1^2-8n+3),\\&c_7=64d_3d_2^2(4d_1^3-18d_1^2+27d_1-9),\\&c_8=-96d_2d_3(2d_1^4-7d_1^3+4d_1^2+6d_1-3),\\&c_9=16d_2^3(2d_1^4-26d_1^3+93d_1^2-117d_1+36),\\&c_{10}=d_1^5-31d_1^4+168d_1^3-360d_1^2+330d_1-90,\\&c_{11}=12d_1^6-134d_1^5+622d_1^4-1484d_1^3+1872d_1^2-1152d_1+252,\\&c_{12}=8(7d_1^5-47d_1^4+121d_1^3-141d_1^2+63d_1-9),\\&c_{13}=16d_1d_2d_3d_4(3d_1^2-8d_1+3),\\&c_{14}=8d_1(d_1^7-4d_1^6-15d_1^5+122d_1^4-287d_1^3+297d_1^2-126d_1+18).\label{72h}\end{split}
\end{align}

\section*{Acknowledgements}

The author KS gratefully acknowledges the travel grant under his CIMPA-ICTP Fellowship.

\end{document}